\documentclass[%
 reprint,
showpacs,
 amsmath,amssymb,
 aps,
]{revtex4-1}

\usepackage{graphicx}% Include figure files
\usepackage{dcolumn}% Align table columns on decimal point
\usepackage{bm}% bold math
\begin{document}

% \preprint{APS/123-QED}

\title{ Photonic Hyper-Crystals}% Force line breaks with \\

\author{Evgenii E. Narimanov}
\affiliation{School of Electrical and Computer Engineering  and Birck Nanotechnology 
Center, \\ Purdue University, West Lafayette, IN 47907, USA
}%

\date{\today}% It is always \today, today,

\begin{abstract}
We introduce a new ``universality class'' of artificial optical media - photonic hyper-crystals. These hyperbolic metamaterials  with periodic spatial variation of  dielectric permittivity on subwavelength scale, combine the features of optical metamaterials and photonic crystals.  In particular, surface waves supported by a hyper-crystal, possess the properties of both the optical Tamm states in photonic crystals and surface plasmon polaritons at the metal-dielectric interface. 
%\begin{description}
% Secondary publications and information retrieval purposes.
%\item[PACS numbers 81.05.Xj, 78.67.Pt]
% May be entered using the \verb+\pacs{#1}+ command.
% You may use the \texttt{description} environment to structure your abstract;
% use the optional argument of the \verb+\item+ command to give the category of each item. 
%\end{description}
\end{abstract}

\maketitle

%\tableofcontents

Metamaterials \cite{ref:metamaterials} and photonic crystals \cite{ref:pc} currently represent the primary building blocks for novel nanophotonic  devices. With the goal of ultimate control over the light propagation, an artificial  optical material must rely on either the effect of  a subwavelength pattern that changes the average electromagnetic response of the medium, \cite{NIM,NIMreview} or on Bragg scattering of light due to a periodic spatial variation that is comparable to the wavelength. \cite{Yablonovich,SJohn} By virtue of this inherent scale separation, the corresponding metamaterial and photonic crystal concepts are generally considered mutually exclusive within the same environment.

\begin{figure}[b]
\includegraphics[width=2.65
   in]{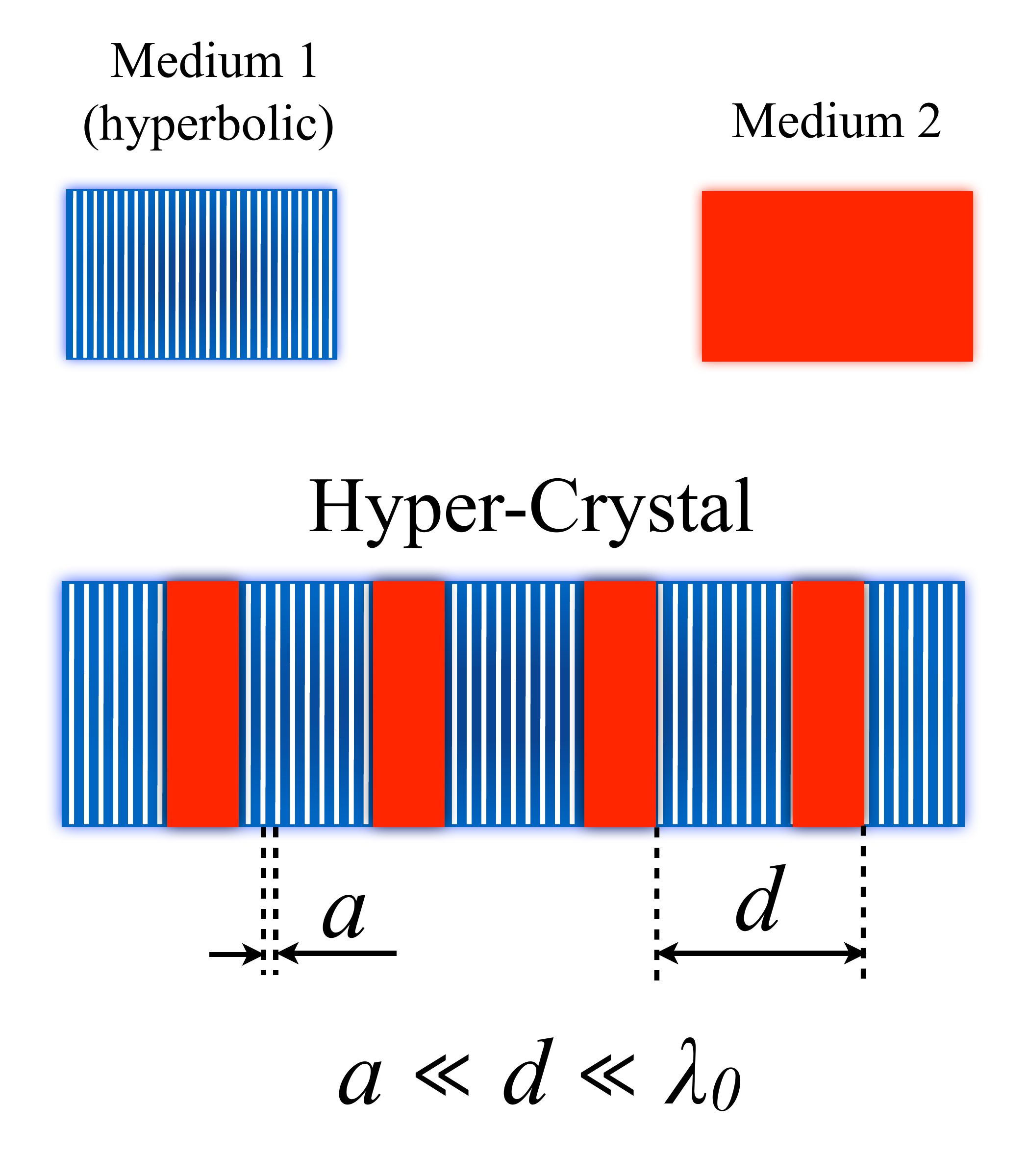}% Here is how to import EPS art
\caption{\label{fig:01} A hyper-crystal is formed by introducing a periodic variation in a hyperbolic medium, with the 
period smaller than the free-space wavelength $d \ll \lambda_0$, but well above the unit cell size of the hyperbolic (meta)material
$d \gg a$. The desired periodic variation can be achieved by introducing a second medium (which could be either a metal, or a dielectric, or another hyperbolic medium with a different dielectric permittivity tensor) in the design on the composite
}
\end{figure}

\begin{figure}[b]
\includegraphics[width=3.5
   in]{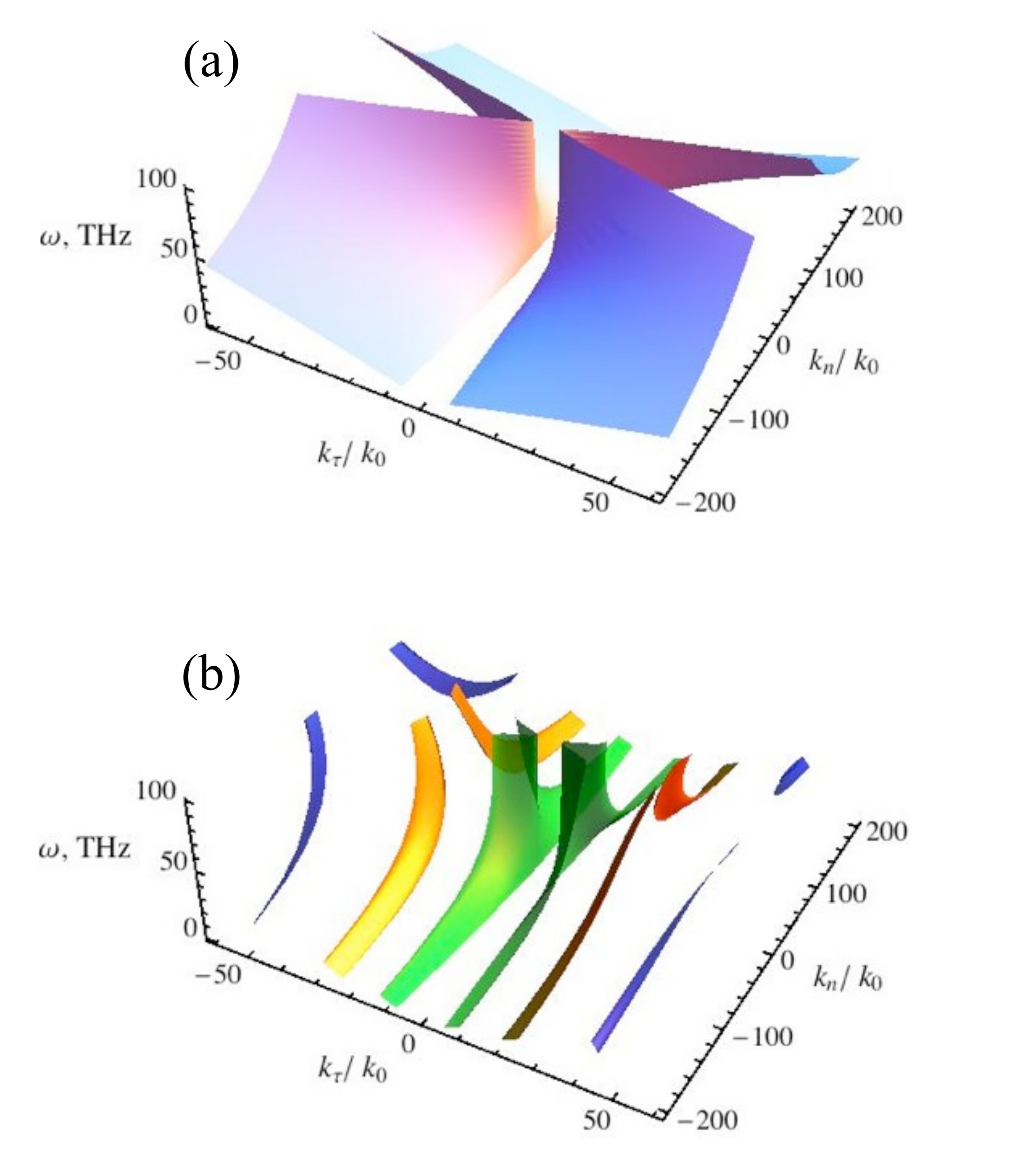}% Here is how to import EPS art
\caption{\label{fig:1}The comparison of the effective medium dispersion of a phonic hyper-crystal (a) to the exact solution (b), in the lossless limit. The hyper-crystal unit cell is formed by $250$ nm of   semiconductor hyperbolic metamaterial  \cite{ref:nmat} ($25$\% $n^+$-doped In$_{0.53}$Ga$_{0.47}$As with $5 \mu$m  plasma wavelength and 
$75$\% Al$_{0.48}$In$_{0.52}$As),  followed by $250$ nm dielectric layer of Al$_{0.48}$In$_{0.52}$As.}
\end{figure}

The situation is however dramatically different in the world of hyperbolic metamaterials \cite{OE2006,SmithPRL}, where
the opposite signs of the dielectric permittivity components in two orthogonal directions ($\epsilon_n \epsilon_\tau < 0$)  lead  to a hyperbolic dispersion of TM-polarized propagating waves
\begin{eqnarray}
\ {k_\tau^2}/ {\epsilon_n} + {k_n^2}/ {\epsilon_\tau}  =  {\omega^2}/{c^2},
\end{eqnarray}
with the wave numbers unlimited by the frequency $\omega$. As a result, a periodic variation in the dielectric permittivity, regardless of how small is its period $d$ (Fig. \ref{fig:01}), will necessarily cause Bragg scattering of these high-$k$ waves, leading to the formation of photonic bandgaps in {\it both} the wavenumber and the frequency domains -- see Fig. \ref{fig:1}. Interestingly, while photonic crystals formed by hyperbolic media have been considered earlier \cite{XiangJOSAB2007,Pan2009}, with the emphasis on omnidirectional band gaps \cite{XiangJOSAB2007} and Goos-H\"anchen shift \cite{Pan2009},  these studies generally focused on the photonic crystal regime $d \sim \lambda_0$ and thus avoided the hyper-crystal limit $d \ll \lambda_0$, where $\lambda_0$ is the corresponding free-space wavelength.

This effect  on the wave propagation and dispersion  by phase space bandgap formation in what is essentially the metamaterial limit,  allows for an unprecedented degree of control of light propagation in photonic hyper-crystals. In particular, 
it can possibly offer  a solution to the loss vs. confinement  conundrum that's been the plague of modern plasmonics. While it was the realization that coupling of photons to charges at metal interfaces allows subdiffraction-limit localization of light that has revived the field of surface plasmons \cite{NaturePhotonicsEditorial2012}, the subwavelength confinement is also the regime of the highest surface plasmon propagation loss -- as it's the photon-electron coupling which is the origin of both of these effects. 

While there is an alternative mechanism of the optical surface state formation  due to Bragg reflection in the band gap of a photonic crystal that is free from high loss, responsible for the so called optical Tamm states \cite{Tamm,TammStates}, it normally does not lead to a subwavength localization. However, in a photonic hyper-crystal where Bragg reflections and associated band gaps persist into the metamaterial limit (see Fig. \ref{fig:1}),  Tamm mechanism is no longer subject to such limitations.  With Bragg reflections taking a part of the ``load" in light confinement,  when compared to the conventional surface plasmon polaritons at the metal-dielectric interface, these ``hyper-plasmons" can therefore show
both stronger localization (larger wave numbers) {\it and} lower loss.

\section{Photonic hyper-crystals: the concept}

Depending on the relation between the optical wavelength $\lambda_0$ and its unit cell size $a$, 
an artificial composite material generally shows two distinct and qualitatively different regimes for wave propagation
and scattering. In the metamaterial limit $a \ll \lambda_0$, \cite{ref:metamaterials} the electromagnetic response of the composite can be 
described in terms of its effective permittivity and permeability tensors, whose elements are defined by the geometry and 
the composition of the unit cell. \cite{NIM} In contrast to this behavior, optical Bragg scattering in the photonic crystal regime $a \gtrsim \lambda_0$ leads a nontrivial wave dispersion that can no longer be described by the averaged refractive index, and to the  formation of the  bandgaps in the propagating waves spectrum.\cite{ref:pc} 

However, this separation into two distinct regimes tacitly assumes that the wavenumbers of the propagating modes supported by the composite, are within the same order of magnitude as the corresponding free-space value. Indeed, in such case the strong inequality $a \ll \lambda_0$ that the phase $k a$ accumulated by the propagating wave across a single cell of a metamaterial, is much smaller than unity, and the effective medium description of the composite is well justified. \cite{Milton}

While generally appropriate in most optical materials, this assumption is violated in the hyperbolic media where the dielectric permittivities in two orthogonal directions directions ($\epsilon_x = \epsilon_y \equiv \epsilon_\tau$ and $\epsilon_z \equiv \epsilon_n$) have opposite signs, leads to the hyperbolic dispersion of TM-polarized propagating waves 
\begin{eqnarray}
k_\tau^2 - \left( - \frac{\epsilon_\tau}{\epsilon_n}\right) k_n^2 & = & \epsilon_\tau \frac{\omega^2}{c^2}
\label{eq:hyperbolic2}
\end{eqnarray}
as shown in Fig. \ref{fig:1}(a) for the case of semiconductor hyperbolic metamaterials based on In$_{0.53}$Ga$_{0.47}$As : Al$_{0.48}$In$_{0.52}$As superlattice \cite{ref:nmat}; note that the optical hyperbolic media can be found in both ``artificial'' \cite{ref:nmat,APB2010} and ``natural'' (e.g. sapphire\cite{ref:sapphire}, bismuth\cite{ref:bismuth-1,ref:bismuth-2,ref:bismuth-3,ref:bismuth-4}, TGS \cite{ref:TGS-1,ref:TGS-2,ref:TGS-3}, and graphite\cite{ref:graphite})  forms. With the validity of Eqn. ({\ref{eq:hyperbolic2}}) only limited by the unit cell size of the medium $a$, a hyperbolic metamaterial generally supports a broad spectrum of high-$k$ waves, leading to the super-singularity in the photonic density of states \cite{PRL}
and wide range of related phenomena -- from enhanced quantum-electrodynamic effects\cite{ZJISEN,NoginovQED,ShalaevQED} to enhanced scattering and reduced reflectivity \cite{stealth}.

As a result, a periodic variation introduced in a hyperbolic medium, even with the period $d \ll \lambda_0$, will lead to strong Bragg scattering of the high-$k$ propagating waves -- despite the fact that the composite remains within the formal bounds of the ``metamaterial limit". Furthermore, comparing to the ``regular'' phonic crystal, the resulting wave dispersion in such ``hyper-crystals'' shows a substantially more complex ``phase diagram'', with multiple allowed and forbidden bands in both frequency and momentum dimensions (see Fig. \ref{fig:1}).

Formally, we define the photonic hyper-crystal regime as the limit
\begin{eqnarray}
a \ll d \ll \lambda_0
\label{eq:hyper-crystals_definition}
\end{eqnarray}
for a photonic crystal with the period $d$ that is formed by (either natural or artificial)  hyperbolic medium with the unit cell size $a$ (see Fig. \ref{fig:01}). With the unit cell size for existing hyperbolic metamaterials on the scale that ranges from $\sim 10$ nm at visible frequencies \cite{APB2010,ScienceTopologicalTransitions,Liu} to $\sim 100$ nm in Mid-IR \cite{ref:nmat},  the  ratio $\lambda_0/a \sim 100$ allows for sufficient ``separation of scales" to satisfy (\ref{eq:hyper-crystals_definition}). Furthermore, with the use of natural hyperbolic materials such as sapphire \cite{ref:sapphire} or bismuth \cite{ref:bismuth-4}, Eqn. (\ref{eq:hyper-crystals_definition})  reduces to simply $d \ll \lambda_0$.

For a hyper-crystal with axial symmetry (such as e.g. a layered composite \cite{ref:nmat}), we can characterize the ``phase space'' of the system by its the dimensionless ``extinction coefficient''
\begin{eqnarray}
\alpha & \equiv & {{\rm Im}\left[ k_n\right] d}
\label{eq:extinction}
\end{eqnarray} 
where $k_n$ is the wavevector component along the optical axis of the composite $z$, so that the intensity decays as $I(z) \propto \exp\left( - 2  \alpha z / d\right)$. In a propagating band, the extinction coefficient is relatively small and entirely defined by the material absorption, with $\alpha = 0$ in the lossless limit. In contrast, within a bandgap,  $\alpha = {\cal O}\left(1\right)$. The propagating bands ``valleys" and band gap ``ridges" in a 3D plot of the extinction coefficient vs. the wavenumber $k_\tau$ and frequency $\omega$ thus allow a straightforward  visualization of the phase space of a hyper-crystal - see Fig. \ref{fig:3}.

\begin{figure}
\includegraphics[width=3.5
   in]{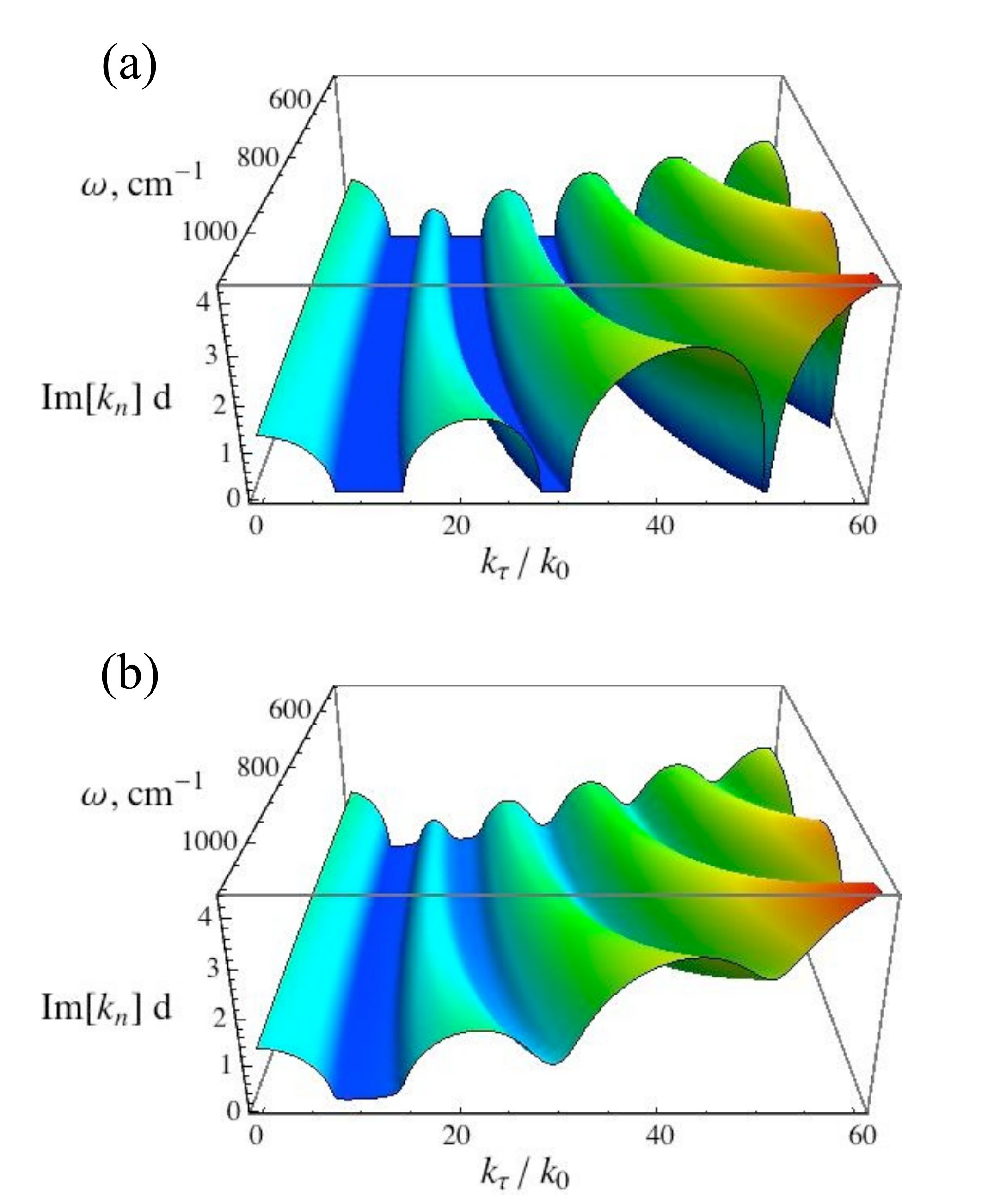}% Here is how to import EPS art
\caption{\label{fig:2} The dimensionless extinction coefficient ${\rm Im}\left[k_n\right] d $  vs. normalized in-plane momentum ($k_\tau / k_0$) and frequency $\omega$, for a semiconductor In$_{0.53}$Ga$_{0.47}$As : Al$_{0.48}$In$_{0.52}$As  / 
In$_{0.53}$Ga$_{0.47}$As  hyper-crystal, calculated in  the  lossless limit (a), and for the actual material losses (b). The single period of the hyper-crystal is defined by $100$ nm layer of $n^+$-doped semiconductor In$_{0.53}$Ga$_{0.47}$As with the plasma wavelength $\lambda_p = 5 \mu$m, followed by $400$ nm thick semiconductor hyperbolic metamaterial. The unit cell   of the hyperbolic metamaterial consists of single layers of the dielectric Al$_{0.48}$In$_{0.52}$As  and $n^+$-doped
 In$_{0.53}$Ga$_{0.47}$As semiconductor (same as in the ``isotropic'' part of the hyper-crystal). 
The dielectric permittivity of   Al$_{0.48}$In$_{0.52}$As is taken to be equal to $10.23$ \cite{ref:nmat}, while the 
  permittivity of $n^+$-doped In$_{0.53}$Ga$_{0.47}$As is described by the Drude model $\epsilon = 12.15 \left(1 - \omega^2 / (\omega^2 + i \omega/\tau)\right)$ \cite{ref:nmat} with the relaxation time $\tau \approx 0.16$ ps, and the plasma frequency $\omega_p$ corresponding to the free-space wavelength of $5 \mu$m.
}
\end{figure}

In Fig. {\ref{fig:2}} we plot the extinction coefficient for the semiconductor superlattice  hyper-crystal ($25$\%  $n^+$-In$_{0.53}$Ga$_{0.47}$As :  $75$\% Al$_{0.48}$In$_{0.52}$As hyperbolic metamaterial unit cell,  with the hyper-crystal  period defined by $250$ nm of hyperbolic metamaterial and $250$ nm of dielectric Al$_{0.48}$In$_{0.52}$As). While multiple 
bandgaps in both frequency and momentum are clearly seen in this plot, note that they are only observed for $k_\tau > k_0$, and therefore are not accessible for light incident from air on a defect-free surface of the hyper-crystal. However, many optical phenomena -- from light scattering \cite{stealth} to near-field radiative thermal transport \cite{PendryThermal,Grefffet1,Greffet2,Zubin,Germans} to coherent thermal radiation \cite{GreffetCoherent} to spontaneous emission \cite{ZJISEN}, are strongly affected by the high-$k$ of the wavenumber spectrum in the system.

\begin{figure}[b]
\includegraphics[width=3.  in]{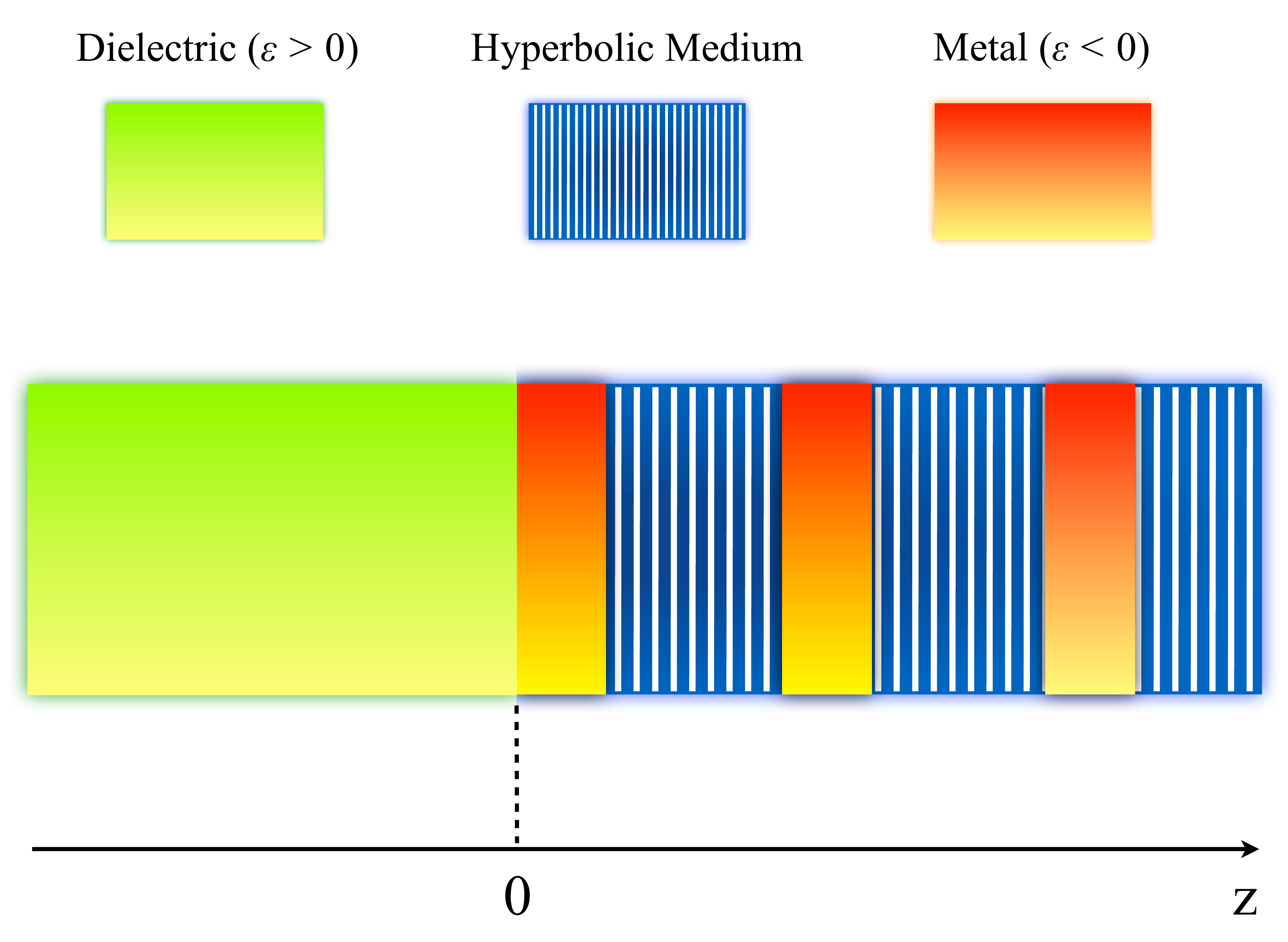}% Here is how to import EPS art
\caption{\label{fig:04} The dielectric -- hyper-crystal interface. }
\end{figure}

The presence of the actual material loss results a nonzero extinction coefficient in the propagation bands, leading to a smaller ``contrast'' between the propagating and  forbidden bands - see Fig. \ref{fig:2}(b). While the extinction coefficient still shows the general ``ridge-valley'' pattern of the ideal lossless hyper-crystal, the typical values of the dimensionless extinction coefficient
(\ref{eq:extinction}) in the high-order bands substantially exceed unity. As a result, a propagating wave from any of such bands will be totally absorbed at the distance of barely a small fraction of the free-space wavelength. Therefore, unless one can substantially reduce the effective loss in the hyper-crystal, a practical application of such volume propagating waves is an entirely loosing preposition.

Of all the existing realizations of hyperbolic media at optical frequencies, the typical ``figure-of-merit'' ${\rm Re}\left[\epsilon\right] /{\rm Im}\left[\epsilon\right] $ for optical hyperbolic media based on the existing plasmonic and polaritonic materials, ranges from unity (e.g. graphene-based hyperbolic metamaterials and graphite) to $\sim 10$ (semiconductor- and silver-based layered hyperbolic media, sapphire) \cite{SashaReview} -- with the only exceptions contributed by the nanowire composites \cite{ref:nanowires} where the relatively low volume fraction of the metal ($\sim 10$\%) results in a proportionally lower loss. \cite{NoginovWires,ZhangWires}  It is this type of the hyperbolic media that holds the most promise for the practical applications of the concept of the hyper-crystal \cite{Marylandhyper-crystals}.

However, in addition to the structured spectrum of the ``bulk'' propagating modes, photonic hyper-crystals also support novel surface waves that combine the features of the regular surface plasmons with those of optical Tamm states. Due to the contribution of Bragg scattering to their formation, even with the actual losses in planar hyper-crystals, these surface waves show both larger wavenumbers  and lower loss than their regular surface plasmon counterparts.

\section{Surface Waves in Photonic hyper-crystals}

We consider TM-polarized waves in a planar hyper-crystal, formed by the layers isotropic material with the permittivity $\epsilon_i$ and uniaxial hyperbolic medium
with the permittivity tensor, 
\begin{eqnarray}
\epsilon & = & 
\left(
\begin{array}{ccc}
\epsilon_\tau & 0 & 0 \\
0 & \epsilon_\tau & 0 \\
0 & 0 & \epsilon_n
\end{array}\right)
\\
\nonumber 
\label{eq:tensor_epsilon}
\end{eqnarray}
with the corresponding thicknesses  $d_i$ and $d_h$ respectively -- see Fig. \ref{fig:04}. The hyper-crystal occupies the half-space $z > 0$ and is terminated at $z = 0$ by the interface with the dielectric medium with the permittivity $\epsilon_d$. This model is exact for natural hyperbolic media, and assumed that $d_h \gg a$ in the case of the hyper-crystal unit cell formed from a hyperbolic metamaterial.

\begin{figure}
\includegraphics[width=3.5
   in]{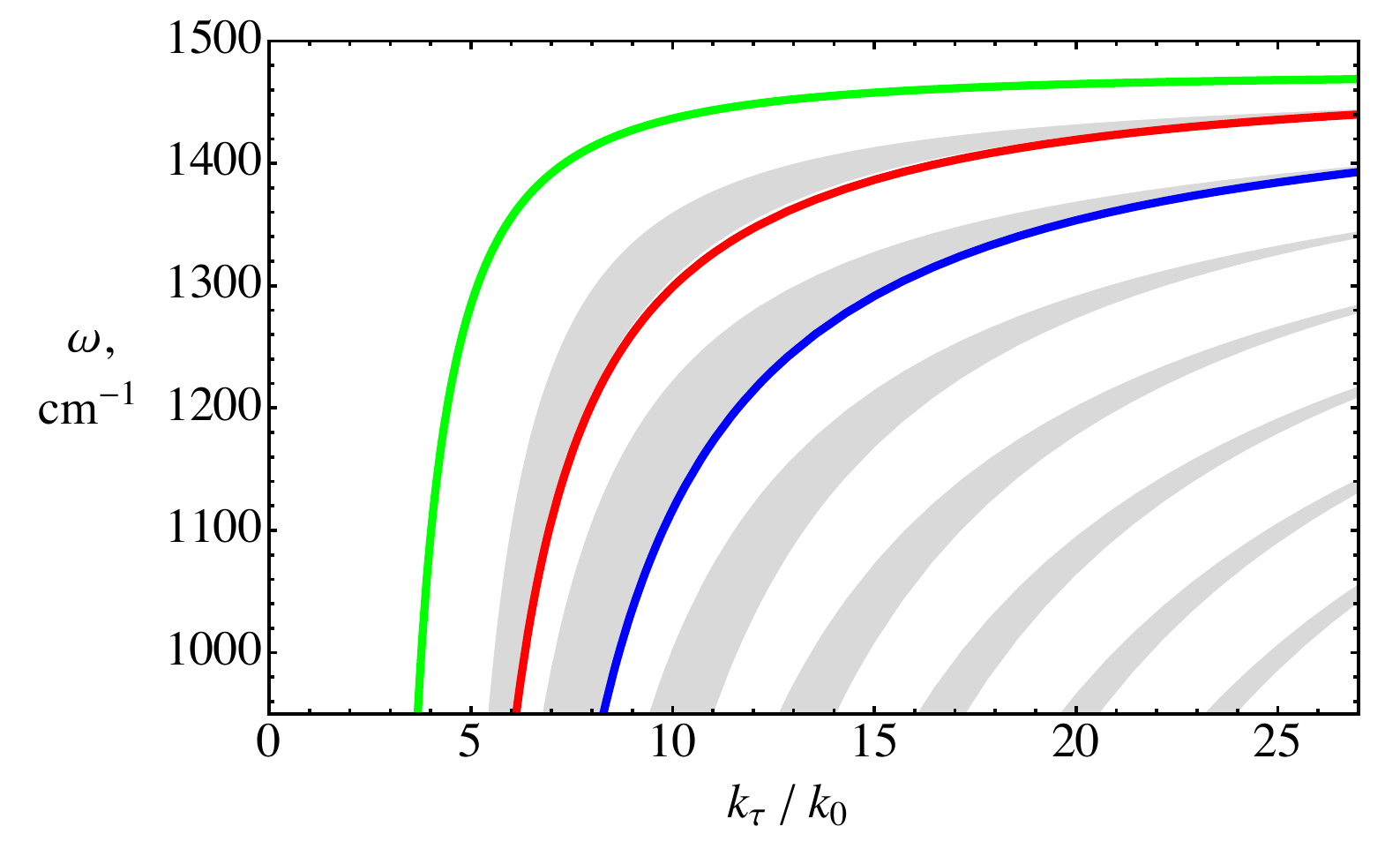}% Here is how to import EPS art
\caption{\label{fig:3}
The surface states at the interface of InGaAs:AlInAs  semiconductor-based hyper-crystal  and dielectric (Al$_{0.48}$In$_{0.52}$As), in the lossless limit. Gray areas  correspond to the volume propagating bands of the hyper crystal. The color code represents surface states of the orders zero (green), one (red) and two (blue).
The unit cell of the hyper-crystal is formed by $1.9 \mu$m-thick semiconductor hyperbolic metamaterial (same as in Fig. \ref{fig:2}), and $100$ nm - wide layer of  $n^+$-doped
 In$_{0.53}$Ga$_{0.47}$As (with the plasma wavelength of $5 \mu$m).
 }
\end{figure}

The calculation of the surface waves in such system is straightforward, following the standard procedure developed for 
1D photonic crystals.\cite{Yeh} We obtain
\begin{eqnarray}
\frac{\epsilon_i}{\epsilon_d} \ 
\sqrt{
\frac{k_\tau^2 - \epsilon_d^2 k_0^2}{k_\tau^2 - \epsilon_i^2 k_0^2}
}
& = &
\frac{{1}/{\lambda} - T_{11} - T_{12}}{1/\lambda - T_{11} + T_{12}}
\label{eq:surfstates1}
\\
\nonumber 
\end{eqnarray}

\noindent
where the $T_{11}$ and $T_{12}$ are the components of the $2 \times 2$ transfer matrix $T$ for the unit cell of the hyper-crystal, \cite{Yeh} and $\lambda$ is its eigenvalue with the absolute value larger  than one. 

Since the determinant of a unit cell transfer matrix in a periodic system is equal to one,
\begin{eqnarray}
{\rm det}\left[ T \right] \equiv T_{11} T_{22} - T_{12} T_{21} = 1,
\end{eqnarray}
Eqn.  (\ref{eq:surfstates1}) can also be expressed as
\begin{eqnarray}
- 
\frac{\epsilon_i}{\epsilon_d}  \ 
\sqrt{
\frac{k_\tau^2 - \epsilon_d^2 k_0^2}{k_\tau^2 - \epsilon_i^2 k_0^2}
}
& = &
\frac{{1}/{\lambda} - T_{22} - T_{21}}{1/\lambda - T_{22} + T_{21}}
\label{eq:surfstates2}
\end{eqnarray}

The system only supports surface states if the trace of the $T$-matrix 
\begin{eqnarray}
\left|\  {\rm Tr} \left[ T \right]\  \right| \equiv \left| T_{11} + T_{22} \right| > 2
\label{eq:Tr2}
\end{eqnarray}
which for ${\rm det}\left[ T\right] = 1$ is a necessary and sufficient condition for the existence of an eigenvalue 
with the modulus $|\lambda | > 1$. 
The physical meaning of this requirement is that the surface states in the hyper-crystal, just as the regular optical Tamm states,  are confined to the bandgaps of the volume propagating modes.

For the transfer matrix of a unit cell of the hyper-crystal and its trace, we obtain:
 \begin{widetext}
\begin{eqnarray}
T_{11} & = & \left[ \cos\left( k_h d_h \right) + \frac{1}{2} \left( \frac{k_h}{\kappa_i} \frac{\epsilon_i}{\epsilon_\tau} -  
\frac{\kappa_i}{k_h}  \frac{\epsilon_\tau}{\epsilon_i} \right) \sin\left(k_h d_h\right) \right] 
% \nonumber \\
% & \times &
 \exp\left( - \kappa_i d_i\right) 
\\
T_{12} & = & \frac{1}{2} \left( \frac{k_h}{\kappa_i} \frac{\epsilon_i}{\epsilon_\tau} +  
\frac{\kappa_i}{k_h}  \frac{\epsilon_\tau}{\epsilon_i} \right) \sin\left(k_h d_h\right) 
 \exp\left(  \kappa_i d_i\right) 
\\
T_{21} & = &  - \frac{1}{2} \left( \frac{k_h}{\kappa_i} \frac{\epsilon_i}{\epsilon_\tau} +  
\frac{\kappa_i}{k_h}  \frac{\epsilon_\tau}{\epsilon_i} \right) \sin\left(k_h d_h\right) 
 \exp\left(  - \kappa_i d_i\right) 
 \\
T_{22} & = & \left[ \cos\left( k_h d_h \right) - \frac{1}{2} \left( \frac{k_h}{\kappa_i} \frac{\epsilon_i}{\epsilon_\tau} -  
\frac{\kappa_i}{k_h}  \frac{\epsilon_\tau}{\epsilon_i} \right) \sin\left(k_h d_h\right) \right] 
%  \nonumber \\
% & \times & 
\exp\left( \kappa_i d_i\right)
 \end{eqnarray}
 \begin{eqnarray}
{\rm Tr}\left[ T\right] & = & 2 \cos\left(k_h d_h\right) \cosh\left(\kappa_i d_i\right) +
\left( \frac{\kappa_i}{k_h}  \frac{\epsilon_\tau}{\epsilon_i}  \right.
% \nonumber \\
% & - & 
-
\left.  \frac{k_h}{\kappa_i} \frac{\epsilon_i}{\epsilon_\tau} 
\right)  \sin\left(k_h d_h\right) \sinh\left(\kappa_i d_i\right) 
\end{eqnarray}
 \end{widetext}
where the wavenumber in the hyperbolic medium
 \begin{eqnarray}
k_h & = & \sqrt{\epsilon_\tau k_0^2 - \left(\epsilon_\tau/\epsilon_n\right) k_\tau^2} 
\end{eqnarray}
and the field decay rate in the isotropic part of the hyper-crystal unit cell 
\begin{eqnarray}
\kappa_i & = & \sqrt{k_\tau^2 - \epsilon_i k_0^2} ,
\end{eqnarray}
while in the dielectric at $z < 0$ 
\begin{eqnarray}
\kappa_d & = & \sqrt{k_\tau^2 - \epsilon_d k_0^2}
\end{eqnarray}
From Eqns. (\ref{eq:surfstates1}) 
\begin{eqnarray}
1/\lambda & = & T_{11} + T_{12} \left(   
2 \  \frac{\epsilon_d}{\epsilon_i} \ 
\sqrt{
\frac{k_\tau^2 - \epsilon_i^2 k_0^2}{k_\tau^2 - \epsilon_d^2 k_0^2}
}
- 1
\right)
\label{eq:1_over_lambda}
\end{eqnarray}
Substituting (\ref{eq:1_over_lambda}) into Eqn. (\ref{eq:surfstates2}), we can 
\begin{widetext}
\begin{eqnarray}
\frac{k_h \epsilon_\tau  \left(\epsilon_d^2 \kappa_i^2 - \epsilon_i^2 \kappa_d^2\right)}{\tan\left( k_h d_h\right)}
-
\frac{\kappa_i \epsilon_i
\left(\epsilon_\tau^2 \kappa_d^2 - \epsilon_i^2 k_h^2\right)}{\tanh\left( \kappa_i d_i\right)}
& = & 
{\kappa_d \epsilon_d} \left(\epsilon_\tau^2 \kappa_i^2 + \epsilon_i^2  k_h^2 \right),
\label{eq:sst}
\end{eqnarray}
\end{widetext}
which can be considered as the fundamental equation for the surface states in a hyper-crystal.
Note that while Eqn. (\ref{eq:sst}) no longer explicitly depends on the eigenvalue $\lambda$, it is still limited to the case of $\left| \lambda \right| > 1$, and should therefore be considered together with the inequality (\ref{eq:Tr2}). 

In the limit $k_\tau \gg k_0 \equiv \omega/c$ we find
\begin{eqnarray}
\frac{A_h}{\tan\left( k_\tau d_* \right)}& = & 1  +  \frac{A_i}{\tanh\left(k_\tau d_i \right)} 
\end{eqnarray}
where 
\begin{eqnarray}
d_* \equiv d_h \sqrt{- \epsilon_\tau / \epsilon_n}
\label{eq:sst_sc}
\end{eqnarray}
 and the coefficients $A_h$ and $A_i$ are defined by the permittivites $\epsilon_\tau$, $\epsilon_n$, $\epsilon_i$ and $\epsilon_d$: 
\begin{eqnarray}
A_h & = &\frac{\epsilon_d}{ \sqrt{ - \epsilon_\tau \epsilon_n}} \ \frac{\epsilon_i^2 - \epsilon_d^2}{\epsilon_i^2 - \epsilon_\tau \epsilon_n} \\
A_i & = & \frac{\epsilon_i}{\epsilon_d} \ \frac{\epsilon_d^2 - \epsilon_\tau \epsilon_n}{\epsilon_i^2 - \epsilon_\tau \epsilon_n} 
\end{eqnarray}
Eqn. (\ref{eq:sst_sc}) should be solved within the limits set by the inequality (\ref{eq:Tr2}), which in this limit ($k_\tau \gg k_0$) reduces to
 \begin{widetext}
\begin{eqnarray}
| \cos\left(k_\tau d_*\right) \cosh\left(k_\tau d_i\right) + 
\frac{1}{2}\left( \sqrt{- \frac{\epsilon_n}{\epsilon_\tau}}\  \frac{\epsilon_\tau}{\epsilon_i} - \sqrt{-\frac{\epsilon_\tau}{\epsilon_n}} \ \frac{\epsilon_i}{\epsilon_\tau}  \right) 
% \nonumber \\
% \nonumber \\
%   \times 
  \sin\left(k_\tau d_*\right) \sinh\left(k_\tau d_i\right) 
| > 1 \ \ \ \ \ \ \ \ \ \ \ \ \ \ \ \ \ 
\end{eqnarray} 
 \end{widetext}

\begin{figure}[t]
\includegraphics[width=3.5
   in]{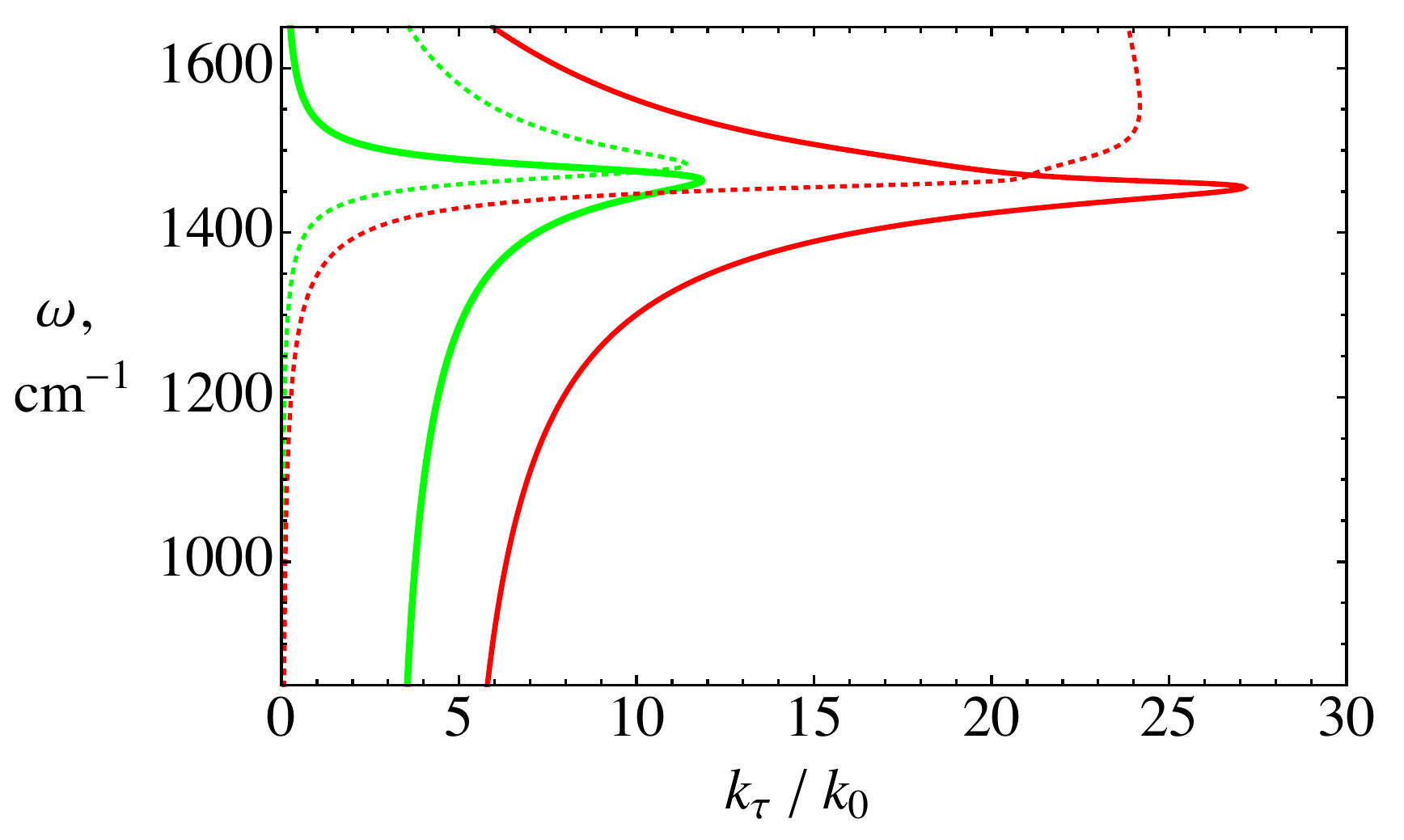}% Here is how to import EPS art
\caption{\label{fig:4} Surface states in the semiconductor hyper-crystal (same as in Fig. \ref{fig:3}), taking into account the actual material absorption, for orders zero (green) and one (red). Solid and dashed lines correspond to respectively the real and the imaginary part of the  in-plane momentum  of the surface modes. }
\end{figure}

In Fig. \ref{fig:3} we plot the first three solutions for the surface states, using the example of the lossless limit for the hyper-crystal formed by semiconductor hyperbolic metamaterial as the hyperbolic medium, and the doped semiconductor (${\rm Re}\left[ \epsilon_i \right] < 0$) as the isotopic material. Note that, aside from the $0$-th order mode whose dispersion is nearly identical to the regular surface plasmon on the interface between the doped semiconductor  and the external dielectric, the surface waves supported by the system, have the frequencies approaching the lower band bounds in the spectrum of the bulk propagating modes of the hyper-crystal.

The presence of the actual material loss limits the maximum values of the surface states wavenumbers, while at the same time extending their existence to the frequencies beyond the maximum values from the lossless case - see Fig. \ref{fig:4}.

\begin{figure}[b]
\includegraphics[width=3.25
   in]{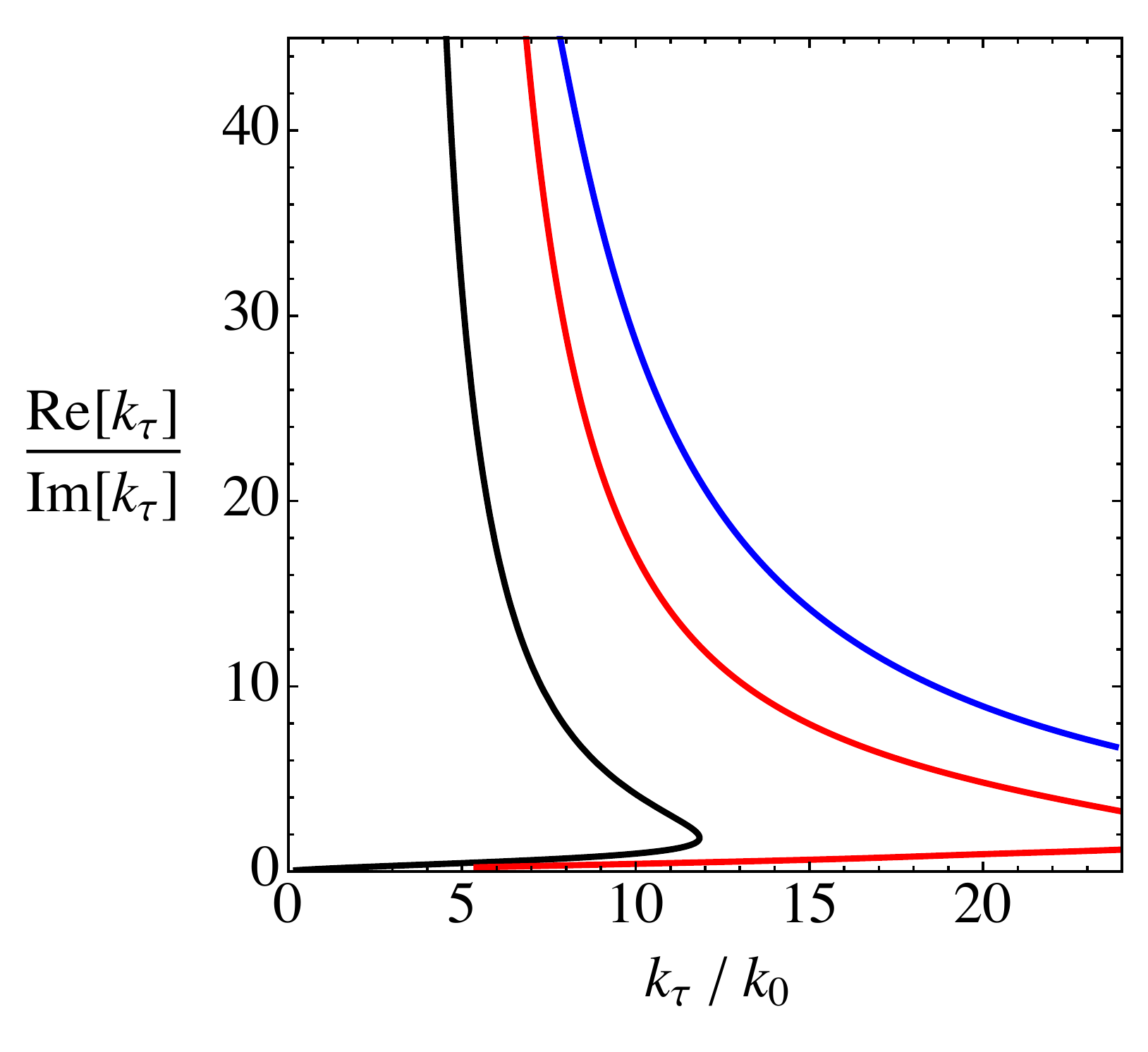}% Here is how to import EPS art
\caption{\label{fig:5} Propagation distance in units of its wavelength (${\rm Re}\left[k_\tau\right] /  {\rm Im}\left[ k_\tau\right]$) vs. its dimensionless wavenumber (normalized to that in free space) of the surface states in the semiconductor hyper-crystal (same as in Figs. \ref{fig:3} and \ref{fig:4}), compared to that of the standard surface plasmon at the metal-dielectric interface (black curve). The surface plasmon corresponds to the interface of $n^+$-doped  
 In$_{0.53}$Ga$_{0.47}$As and dielectric Al$_{0.48}$In$_{0.52}$As \cite{ref:nmat}  -- the same materials as those that form the  hyper-crystal.}
\end{figure}

The limits on the wavenumbers and the propagation distance of the surface waves in hyper-crystals are illustrated in 
Fig. \ref{fig:5},
where we plot the standard  ``figure-or-merit" \cite{ref:figure-of-merit} ${\rm Re}\left[k\right]/ \  {\rm Im}\left[k\right]$ proportional to the propagation distance in units of the actual wavelength, vs. the confinement / wavelength compression factor $k_\tau / k_0$. There, the first- and second-order modes (red and blue curves respectively) are compared to the regular surface plasmon (black line) at the interface of the same materials as those forming the hyper-crystal (doped  InGaAs as the ``metal"  and AlInAs as the dielectric). Note that, with the same material loss, these ``hyper-plasmons'' show more than an  { order-of-magnitude} improvement in propagation distance at the same value of $k_\tau$, and more than a factor of three enhancement of maximum accessible wavenumber.

\section{Conclusions}

Photonic hyper-crystal, the unifying concept of until recently mutually exclusive limits of metamaterials and photonic crystals, dramatically extends the accessible ``phase space'' in electromagnetic material response, and has the potential to find
many applications in nanophotonics.   While the hyper-crystal represents an extra level in the device complexity, its fabrication does not require anything beyond the standard methods used in metamaterials community \cite{ref:metamaterials,ref:nmat}.  

This work was partially supported by NSF
Center for Photonic and Multiscale Nanomaterials,  ARO MURI and Gordon
and Betty Moore Foundation.

\end{document}